\documentclass[preprint,review,12pt]{elsarticle}

\journal{Information and Software Technology}
\usepackage{xcolor}
\usepackage{subcaption}
\usepackage{tabularx}
\usepackage{multirow}
\usepackage{longtable}
\usepackage{afterpage}
\usepackage{hyperref} % Load hyperref first
\usepackage{booktabs}
\usepackage{enumitem}
\usepackage{tikz}
\usetikzlibrary{positioning}
\usepackage{fontawesome5}
\usepackage{pdflscape}
\usepackage{verbatim}

\definecolor{darkgreen}{rgb}{0.0, 0.0, 0.0} 
\definecolor{darkblue}{rgb}{0.0, 0.0, 0.0} 

\newcommand{\todo}[1]{}
\renewcommand{\todo}[1]{{\color{red} TODO: {#1}}}
\def\mybarhhigh#1#2{%%
   {\color{black}\rule{#1mm}{5pt}}  #2}

\usepackage[]{mdframed}
\definecolor{steBoxLine}{rgb}{0.0, 0.0, 0.0}
\mdfdefinestyle{stebox2}{%
linecolor=steBoxLine,linewidth=2pt,%
leftmargin=0cm,rightmargin=0cm,%
roundcorner=5pt,
topline=false,bottomline=false,rightline=true,leftline=true,%
backgroundcolor=gray!05
}

\begin{document}
\begin{frontmatter}

%% Title, authors and addresses

%% use the tnoteref command within \title for footnotes;
%% use the tnotetext command for theassociated footnote;
%% use the fnref command within \author or \address for footnotes;
%% use the fntext command for theassociated footnote;
%% use the corref command within \author for corresponding author footnotes;
%% use the cortext command for theassociated footnote;
%% use the ead command for the email address,
%% and the form \ead[url] for the home page:
%% \title{Title\tnoteref{label1}}
%% \tnotetext[label1]{}
%% \author{Name\corref{cor1}\fnref{label2}}
%% \ead{email address}
%% \ead[url]{home page}
%% \fntext[label2]{}
%% \cortext[cor1]{}
%% \affiliation{organization={},
%%             addressline={},
%%             city={},
%%             postcode={},
%%             state={},
%%             country={}}
%% \fntext[label3]{}

\title{User-Centric Requirements Prioritization in mHealth Applications: Insights from a Discrete Choice Experiment}

%% use optional labels to link authors explicitly to addresses:
%% \author[label1,label2]{}
%% \affiliation[label1]{organization={},
%%             addressline={},
%%             city={},
%%             postcode={},
%%             state={},
%%             country={}}
%%
%% \affiliation[label2]{organization={},
%%             addressline={},
%%             city={},
%%             postcode={},
%%             state={},
%%             country={}}

\author[Monash]{Wei Wang\corref{cor1}}
\cortext[cor1]{Corresponding author}
\ead{wei.wang5@monash.edu}

\author[Deakin]{Hourieh Khalajzadeh}
\ead{hkhalajzadeh@deakin.edu.au }

\author[Monash]{John Grundy}
\ead{john.grundy@monash.edu}

\author[Deakin]{Anuradha Madugalla}
\ead{anuradha.madugalla@deakin.edu.au}

\author[Swinburne]{Humphrey O. Obie}
\ead{hobie@swin.edu.au}

\affiliation[Monash]{organization={Monash University},%Department and Organization
            % addressline={}, 
            city={Melbourne},
            % postcode={}, 
            % state={},
            country={Australia}}

\affiliation[Deakin]{organization={Deakin University},%Department and Organization
            % addressline={}, 
            city={Melbourne},
            % postcode={}, 
            % state={},
            country={Australia}}

% \affiliation[UniMelb]{organization={University of Melbourne},%Department and Organization
%             % addressline={}, 
%             city={Melbourne},
%             % postcode={}, 
%             % state={},
%             country={Australia}}
\affiliation[Swinburne]{organization={Swinburne University of Technology},%Department and Organization
            % addressline={}, 
            city={Melbourne},
            % postcode={}, 
            % state={},
            country={Australia}}

% \author{Wei Wang\fnref{First}\corref{cor1}}
% \author{Hourieh Khalajzadeh\fnref{Second}}
% \author{John Grundy\fnref{First}}
% \author{Anuradha Madugalla\fnref{First}}
% \author{Jenny McIntosh\fnref{Third}
% \author{Humphrey O. Obie\fnref{First}}

% }

% \cortext[cor1]{Corresponding Author}

% \affiliation [First]{organization={Monash University},%Department and Organization
%             addressline={Wellington Rd}, 
%             city{Clayton}
%             state{Victoria}
%             country{Australia}
%             postcode{3031}}
% \affiliation [Second]{organization={Deakin University},%Department and Organization
%             addressline={221 Burwood Hwy}, 
%             city{Burwood}
%             state{Victoria}
%             country{Australia}
%             postcode{3125}}            
% \affiliation [Third]{organization={The University of Melbourne},%Department and Organization
%             addressline={Parkville Grattan Street}, 
%             city{Parkville}
%             state{Victoria}
%             country{Australia}
%             postcode{3010}}   
            
\begin{abstract}

\textbf{Context:} Mobile health (mHealth) applications are widely used for chronic disease management, but usability and accessibility challenges persist due to the diverse needs of users. Adaptive User Interfaces (AUIs) offer a promising approach to personalizing interactions and improving user experience. However, their adoption remains limited, partly due to a lack of understanding of how users perceive and evaluate different adaptation strategies. Addressing this gap is crucial for advancing user-centered design and requirements engineering in software systems for health contexts.

\noindent\textbf{Objective:} This study identifies key factors influencing user preferences and trade-offs in mHealth adaptation design.

\noindent\textbf{Method:} A Discrete Choice Experiment (DCE) was conducted with 186 participants living with chronic conditions who regularly use mHealth applications. Each participant completed a series of choice tasks, selecting their preferred adaptation designs from scenarios composed of six attributes with varying levels. A mixed logit model was applied to examine preference heterogeneity. Subgroup analyses were also conducted to explore variations in preferences across age, gender, health condition, and coping mechanism.

\noindent\textbf{Results:} Participants preferred adaptation designs that preserved usability, offered controllability, introduced changes infrequently, and applied small-scale modifications. Conversely, adaptations affecting frequently used functions and those involving caregiver input were generally viewed less favorably. These findings highlight key trade-offs that influence user acceptance of adaptive mHealth interfaces.

\noindent\textbf{Conclusion:} This study employs a data-driven approach to quantify user preferences, identify key trade-offs, and reveal variations across demographic and behavioral subgroups through preference heterogeneity modeling. These insights provide actionable guidance for designing more user-centered adaptive interfaces and contribute to advancing requirements prioritization practices in software engineering—particularly in the context of health technologies.

\end{abstract}

\end{frontmatter}

\section{Introduction}
Chronic diseases such as diabetes, heart disease, stroke, and cancer remain leading causes of death worldwide. In 2022, they accounted for nearly 40\% of all U.S. deaths and significantly increased COVID-19 risks \cite{hacker2024burden,ahmad2023provisional,People_2024}. Effective chronic disease management relies on sustained adherence to care plans, essential for better health outcomes, quality of life, and cost-efficiency \cite{hamine2015impact}. \textbf{Mobile health (mHealth)} applications provide crucial features like symptom tracking, medication reminders, remote monitoring, and telehealth, which aids in disease management through smartphones and wearables \cite{bricca2022quality}. mHealth applications should enhance \textit{accessibility and flexibility} to effectively engage users with chronic diseases \cite{choe2017semi}. Most mHealth applications for chronic disease management use a\textit{ one-size-fits-all} design approach that doesn't meet users' diverse needs \cite{grua2020reference,reiners2019sociodemographic}. Patients with chronic diseases differ in backgrounds, expertise, and demographic, psychological, and cognitive traits \cite{Vasilyeva2005}. As chronic diseases progress, patients' needs change, indicating diverse requirements \textit{ differ not only between individuals but also throughout their lifetime} \cite{di2019chronic}. These long-term conditions \cite{harvey2012future} often come with \textit{comorbid medical and psychopathological conditions }\cite{di2019chronic}, increasing user needs and functional requirements \cite{gregor2002designing}. \color{darkgreen} However, few studies have systematically addressed how to design mHealth systems that can adapt to these complex, dynamic, and individualized user needs over time.

\color{black}
Smartphones and tablets provide flexible access to health information \cite{Vogt2010}, while wireless sensors and location systems improve contextual data collection \cite{Vogt2010}. However, these technologies create variability in user environments and needs, making fixed \textbf{User Interfaces (UIs)} inadequate for many mHealth applications, requiring more adaptable solutions to serve diverse users \cite{wang2024adaptive}. Several studies have been conducted on the creation of \textbf{Adaptive User Interfaces (AUIs)} for mHealth applications for various chronic diseases \cite{burke2009optimising, pagiatakis2020intelligent, mohan2008medinet, awada2018adaptive, jabeen2019improving}. Many studies on AUIs have involved minimal user engagement during the development and evaluation phases, making it challenging to assess user preferences for these adaptations \cite{wang2024adaptive}. However, research has consistently shown a trade-off between adaptive mechanisms and usability \cite{paymans2004usability, gajos2008predictability, wesson2010can, lavie2010benefits, peissner2013user, wang2024adaptiveicse, wang2024designing}. These trade-offs complicate UI adaptation decisions, as users differ in prioritizing goals, competence, app usage, and scenarios \cite{wang2024designing}. Limited research has focused on understanding user priorities in adaptive UI design, with existing studies often constrained to specific tasks, applications, and small participant samples \cite{lavie2010benefits,Gajos2017TheIO}. While \textit{requirement prioritization} is a critical aspect of software development \cite{achimugu2014systematic, riegel2015systematic}, traditional approaches tend to focus on stakeholder demands, often overlooking the nuanced user preferences and trade-offs essential for creating truly user-centric designs. Bridging this gap requires systematic methods that capture how users weigh competing design factors in real-world contexts. \color{darkgreen} Our study contributes to this gap by using a large-scale Discrete Choice Experiment to empirically model user trade-offs and preferences in adaptive mHealth UI design across diverse user groups.
\color{black}

\textbf{Discrete Choice Experiments (DCEs)} offer a promising yet underexplored approach in \textbf{Software Engineering (SE)} by providing a data-driven, structured method for quantifying user preferences and modeling trade-offs between competing requirements. DCE is a survey-based technique that presents participants with hypothetical choice scenarios to select their preferred option, allowing researchers to infer the relative importance of different attributes and understand how users prioritize competing factors in decision-making \cite{johnson2013constructing}. Although DCEs have been widely applied in healthcare and economics to tackle complex decision-making challenges \cite{de2012discrete, tunnessen2020patients, jiang2023patient,al2023patient,jonker2020covid,nittas2020self,simblett2023patient}, their potential to improve requirements prioritization in software development remains largely unexplored. This study seeks to address this gap by investigating the key factors influencing AUI adoption in mHealth applications for chronic disease management through DCEs. \color{darkblue}Our research makes several \textbf{key contributions} to the field of adaptive mHealth design. First, it provides a data-driven approach to quantifying user preferences and assessing the relative importance of key adaptation attributes in mHealth applications. Second, it uncovers important trade-offs between competing design factors and reveals preference heterogeneity across user subgroups based on demographic and behavioral characteristics. Finally, it lays the foundation for the development of next-generation adaptive mHealth applications and identifies future research directions within the broader field of software engineering.
% Our DCE study with chronic disease app designs and users indicates that while usability is the most crucial attribute across all subgroups, controllability, frequency, and granularity of adaptations also hold importance but vary by subgroup. Users might resist adaptation to frequently used features and see caregiver involvement as decreasing the need for adaptations.
\color{black}
The rest of the paper is organized as follows. Section \ref{sec:relatedwork} reviews the related work, focusing on the trade-offs between different adaptation factors and the application of DCE in healthcare research and software engineering research. Section \ref{sec:method} outlines the research methodology used to conduct the DCE study. The results are presented in Section \ref{sec:results}, followed by Section \ref{sec:discussion}, which explores key implications for mHealth developers and researchers. Sections \ref{sec:threat} and \ref{sec:conclusion} conclude by outlining the study's limitations, contributions, and offering suggestions for future research directions.

\section{Related work}\label{sec:relatedwork}

\begin{figure}[!ht]
     \begin{subfigure}[b]{0.31\textwidth}
         \centering
         \includegraphics[width=\textwidth]{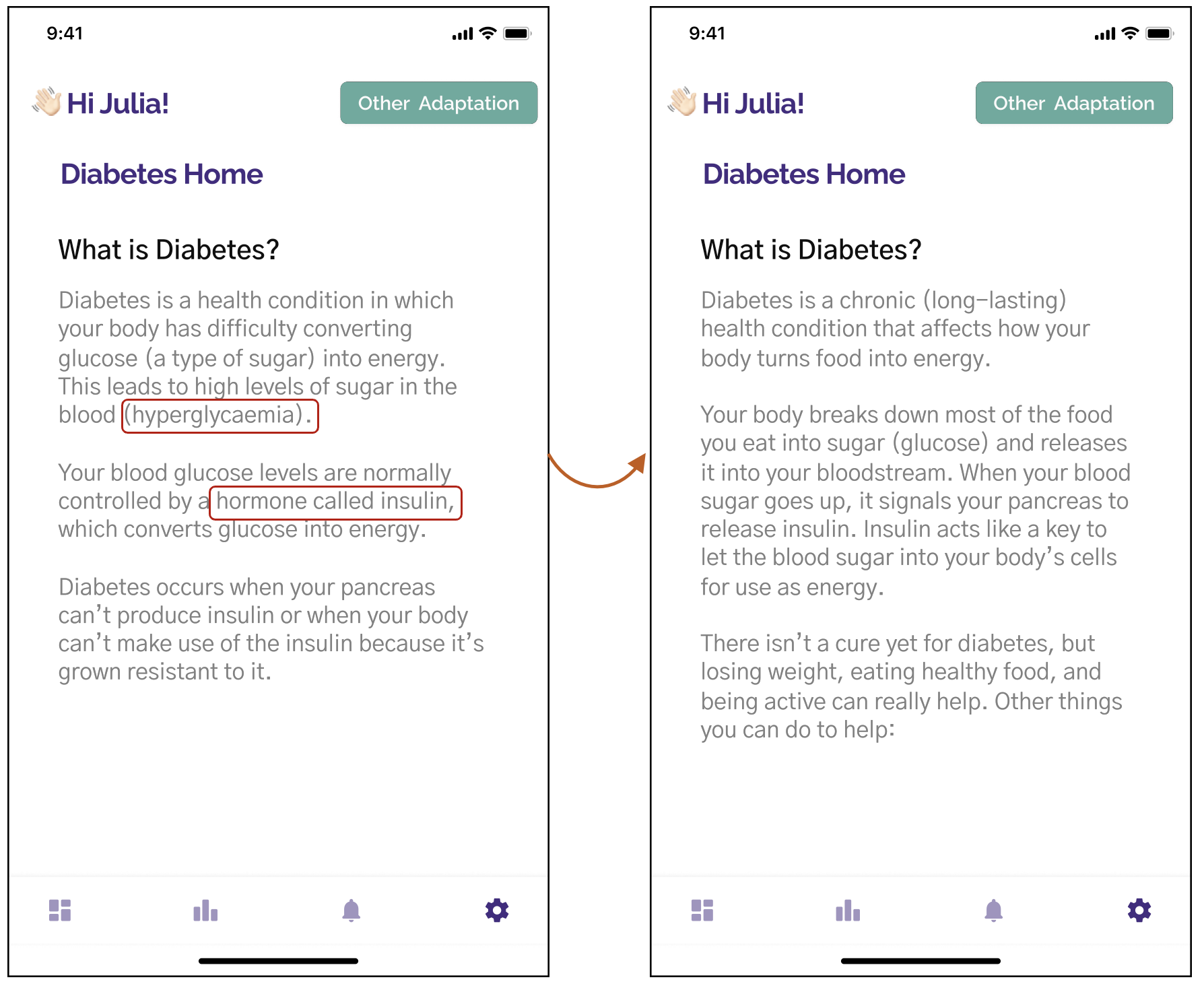}
         \caption{\scriptsize Easy-to-understand language}
         \label{(sub:preinf)}
     \end{subfigure}
     \hfill
     \begin{subfigure}[b]{0.31\textwidth}
         \centering
         \includegraphics[width=0.82\textwidth]{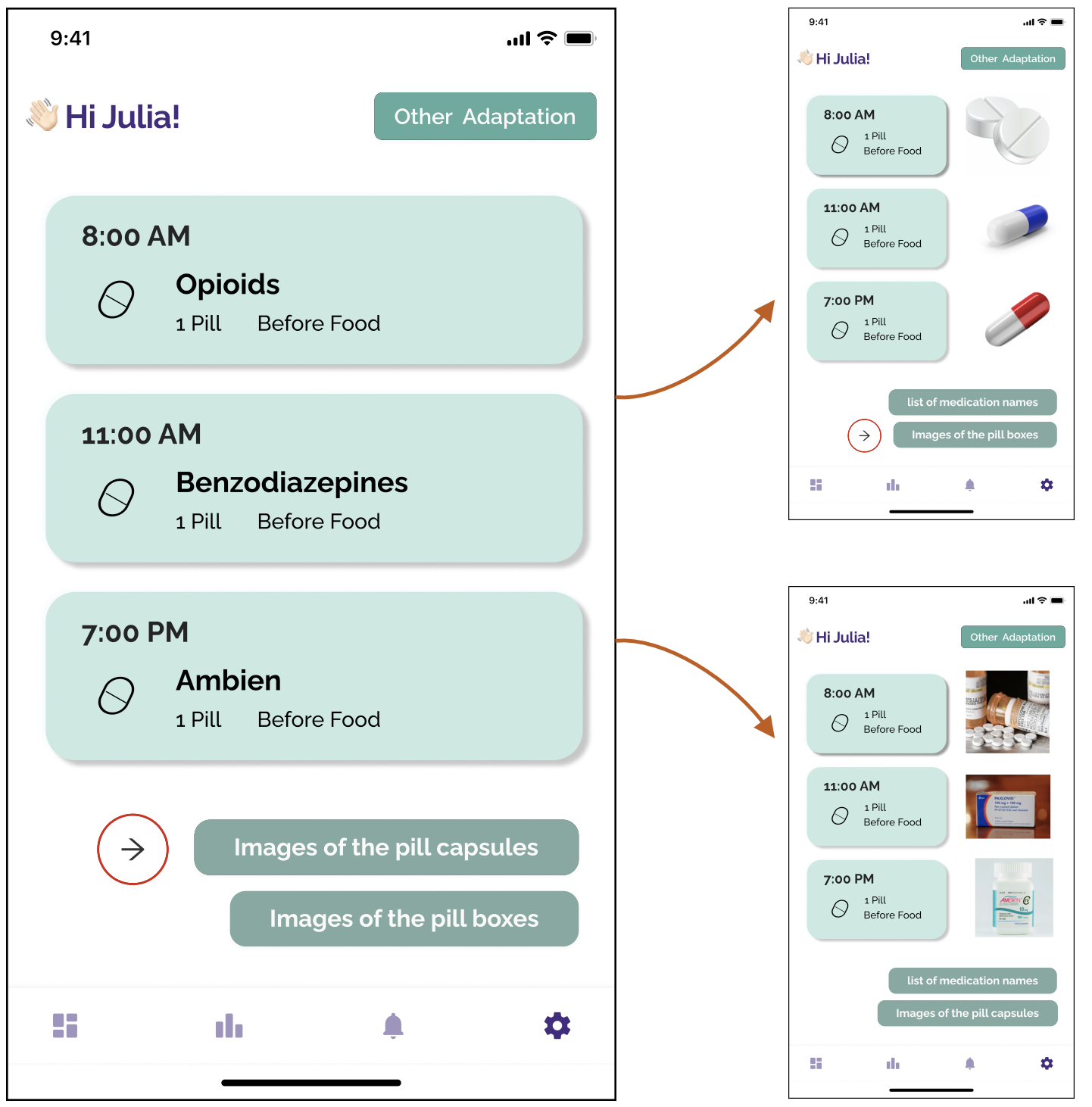}
         \caption{\scriptsize Text-to-image conversion}
         \label{(sub:conima)}
     \end{subfigure}
     \hfill
          \begin{subfigure}[b]{0.35\textwidth}
         \centering
         \includegraphics[width=0.9\textwidth]{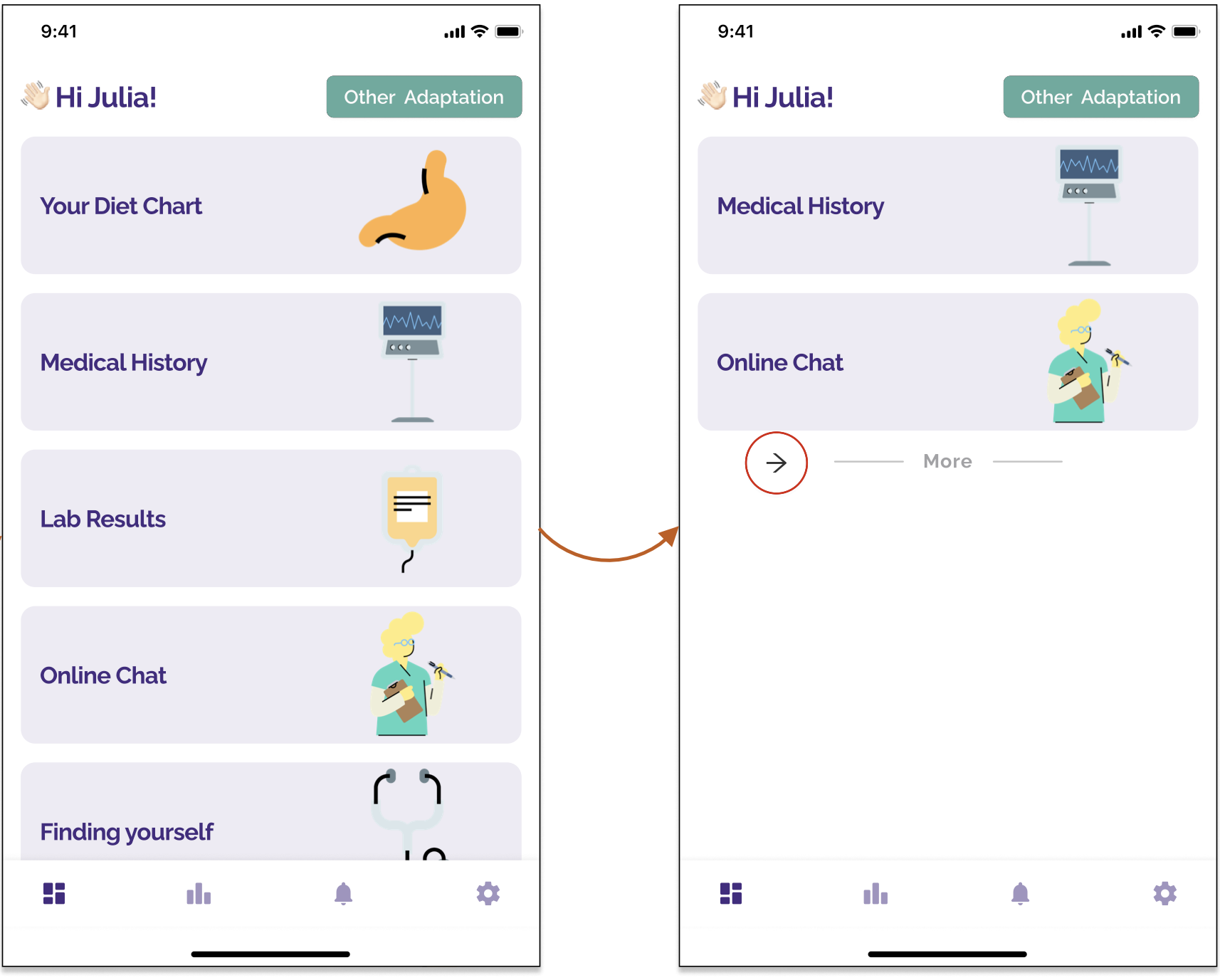}
         \caption{\scriptsize Interface elements rearrangement}
         \label{(sub:conele)}
     \end{subfigure}

    \caption{Examples of prototype of content adaptation.}

\end{figure}
\subsection{Adaptive User Interfaces for mHealth applications}
AUIs have been deployed across a spectrum of mHealth applications, ranging from stroke rehabilitation \cite{burke2009optimising}, diabetes \cite{pagiatakis2020intelligent}, cardiac disease \cite{mohan2008medinet}, dementia \cite{awada2018adaptive} and Parkinson's disease \cite{jabeen2019improving}. A systematic literature review (SLR) \cite{wang2024adaptive} identified three primary types of Adaptive User Interfaces (AUIs) in mHealth applications: presentation adaptation, content adaptation, and behavior adaptation. \textit{Presentation adaptation} focuses on improving user experience by modifying the visual aspects of the UI, such as adjusting colors, positioning, and font sizes to \textit{better align with user preferences and accessibility needs. Content adaptation} tailors the information displayed to better match individual user needs. For example, this can include simplified text representation that enhances readability through clear and concise language (see Figure \ref{(sub:preinf)}), text-to-image conversion that transforms complex textual content into visual formats for better understanding (see Figure \ref{(sub:conima)}), and interface element rearrangement (see Figure \ref{(sub:conele)}). \textit{Behavior adaptation} is a more complex form of adaptation that spans multiple aspects of content and presentation while also modifying system interactions. For example, this can involve adjusting navigation permissions, customizing motivation techniques to encourage behavior change according to user type or health status, and modifying interface modalities to accommodate different usage contexts. These mechanisms enable mHealth apps to deliver personalized and accessible experiences that align with user needs.

\subsection{Trade-offs between different UI adaptation factors}

Existing research on AUI for mHealth often lacks detailed reporting on the design and development process, particularly in the early stages that involve user input, making it difficult to align user preferences and requirements with the final AUI solutions \cite{wang2024adaptive}. In addition, many evaluations assess the overall effectiveness of applications rather than specific adaptive features, often using small participant sample, which limits the generalizability of findings on AUI effectiveness \cite{wang2024adaptive}. As a result, the limited involvement of users in both the development and evaluation of AUIs raises uncertainty about whether users truly prefer adaptive interfaces as a whole or specific adaptive functionalities within them. In addition, research has identified distinct trade-offs between adaptation features and design choices. Adaptations can improve or alter usability \cite{paymans2004usability,wesson2010can,wang2024adaptiveicse}, and may introduce predictability challenges, making it difficult for users to anticipate subsequent actions \cite{gajos2008predictability}. The trade-offs associated with adaptation are influenced by the balance between routine and non-routine tasks, with older individuals experiencing greater adaptation benefits than their younger counterparts \cite{lavie2010benefits}. \citet{peissner2013user} investigated the balance between adaptability and user control by analyzing situations that require explicit user confirmation prior to adaptation and those that do not. These studies are restricted by \textit{a limited number of participants} and a specific demographic profile of users, such as older participants, and focus on specific adaptation conditions. The Technology Acceptance Model (TAM) \cite{davis1989technology} and the Unified Theory of Acceptance and Use of Technology (UTAUT) \cite{venkatesh2012consumer} offer a lens for understanding these trade-offs: adaptive features can enhance perceived usefulness and ease of use if they align with user goals, but unpredictability or poor transparency may hinder adoption. UTAUT also highlights the role of facilitating conditions (e.g., customization options). However, it is still unclear how users’ implicit preferences align with these trade-offs, the level of control they expect over system-driven decisions, and how priority variations differ between user groups, as only a few studies have explored these aspects. Although advances in AI allow systems to learn and adapt to human behavior dynamically, the lack of transparency in how such systems prioritize features raises critical questions \cite{ehsan2021expanding}. Understanding these priorities is crucial to ensure that AI systems adapt in ways that align with user expectations, improve usability, and build trust.

\subsection{Requirements prioritization}

% Requirements elicitation is one of the most critical activities in requirements engineering (RE), which, in turn, is a major determinant of successful development of information systems \cite{pohl2010fundamentals}. In conventional RE, requirements
% are elicited from domain knowledge obtained from stakeholders, relying primarily on qualitative data collection methods (e.g., interviews, workshops, and focus group discussions) \cite{pacheco2018requirements}. 
Requirements prioritization is a crucial aspect of software development, particularly in mobile application development, where frequent system updates, limited resources, response to user feedback, and challenges such as decreased user adherence due to overwhelming choices complicate the process \cite{jantunen2011challenge}. Given the constraints of human resources, technical feasibility, cost, and schedule, it is rarely feasible to implement all the identified requirements. Therefore, the primary goal of requirements prioritization is to select an optimal set of requirements that balances stakeholder demands with available resources while effectively addressing trade-offs \cite{pitangueira2015software}. \color{darkblue}Several SLRs have examined requirements prioritization techniques \cite{achimugu2014systematic, riegel2015systematic}; these works provide a comprehensive overview of existing methods and are complementary to our study. However, most of the approaches they review primarily emphasize the perspectives of \textit{project stakeholders} rather than directly capture end-user preferences. This limitation is particularly evident in user-centric software systems, where conflicts and trade-offs between usability, functionality, and other quality attributes are common \cite{johnson2020designing}. In contrast, our work foregrounds end-user preference elicitation as a core input to requirements prioritization.\color{black}

\subsection{Discrete choice experiments}

\textit{Stated preference methods} are survey-based approaches designed to elicit people's preferences towards specific, often poorly understood behaviors, with DCE being the most widely used type \cite{johnston2017contemporary, johnson2013constructing}. In recent years, DCEs have become accepted as a useful method for \textbf{quantifying societal preferences} and \textbf{setting priorities} in healthcare by allowing an understanding of the relative importance of attributes affecting healthcare decisions and especially the trade-offs that individuals are willing to make between these attributes \cite{de2012discrete, soekhai2019discrete}. Numerous DCE studies have been reported that address patient experiences and health outcomes \cite{de2012discrete}. Some research explores optimal methods for providing health services or medical treatments. For example, certain studies focus on patients' treatment preferences for depressive and anxiety disorders \cite{tunnessen2020patients}. A study investigates how patients weigh their options among different attributes of targeted pharmacotherapy, such as cost, effectiveness, and side effects \cite{jiang2023patient}, and other studies are dedicated to the treatment of epilepsy \cite{al2023patient}. 

The development of mHealth technologies that improve access to health knowledge and information is crucial to address physical and social inequities \cite{han2010professional}, particularly given the high levels of non-adherence to treatment regimens in developing nations \cite{beaglehole2008improving, alwan2009review}. However, the implementation of mHealth applications presents significant challenges, with mixed results in their effectiveness in supporting the adherence to treatment plans \cite{hamine2015impact, gandapur2016role}. Another key issue is the low adoption and high abandonment rates among users \cite{greenhalgh2017beyond}, underscoring a persistent gap between the design and consumer preferences \cite{hamine2015impact}. Few studies have explored strategies for improving mHealth technology design, including COVID-19 contact tracing applications \cite{jonker2020covid} and self-monitoring tools that track sun exposure duration and intensity \cite{nittas2020self}. Additionally, research has examined patient preferences for key factors driving the adoption of mHealth solutions for mental health management, such as those that address depression \cite{simblett2023patient}. 

Traditional requirements elicitation methods often fail to capture the \textbf{complexities} of user decision-making and trade-offs. Although all requirements are typically considered essential, research shows that their \textbf{importance varies} between stakeholders \cite{achimugu2014systematic}. DCEs offer a solution by \textit{simulating real-world decision-making}, requiring users to weigh multiple trade-offs simultaneously rather than evaluating features in isolation \cite{johnston2017contemporary, johnson2013constructing}. Unlike standard surveys, which rank or rate attributes independently, DCEs present choice scenarios with varying attribute combinations, enabling researchers to quantify user preferences and support more nuanced decision-making \cite{brett2013quantifying}. Integrating DCE into software development presents a novel opportunity to align mHealth features with user expectations and enhance prioritization effectiveness. However, despite its proven benefits in other fields \cite{de2012discrete}, DCEs have not yet been widely adopted in SE, highlighting a critical gap in using this method for informed design and decision-making in this field.

 % For example, research has successfully applied DCE to explore consumer preferences for purchasing mHealth applications \cite{xie2023consumers}.
\section{Method}\label{sec:method}

We conducted a DCE survey study to explore the perceptions of individuals with chronic diseases and prior experience using mHealth applications on various aspects of adaptation. \color{darkgreen} Given the need for a focused and interpretable approach in a short communication format, the DCE method offers a concise yet powerful way to quantify user preferences through structured choice tasks. \color{black}In the survey, participants were repeatedly presented with \textbf{two} versions of the same mHealth application and asked to select their preferred option (see the example question in Figure \ref{fig:discreteexample}). By analyzing these repeated choices, we can statistically determine which attributes—and their specific levels—participants value most, providing information on factors that maximize user utility \cite{ryan2001use, ryan2000using}.

Our methodology follows the guidelines reported by the ISPOR Good Research Practices for Conjoint Analysis Task Force \cite{bridges2011conjoint} and is structured around five steps: (1) Key attributes identification, (2) Survey development, (3) Piloting and survey adjustments, (4) Data collection, and (5) Choice data analysis. The survey was administered online and received ethical approval from the Monash University Human Research Ethics Committee (Project ID: 40245).

\subsection{Key attributes identification}\label{sec: attribute}
% Requires: \usepackage{graphicx}
\begin{table}[h]
\renewcommand{\arraystretch}{0.8}
    \centering
    \caption{Identified attributes and levels} 
    \resizebox{0.98\textwidth}{!}{%
    \begin{tabular}{p{57mm}p{3mm}p{135mm}}
        \hline
        \textbf{Attribute} & \textbf{\#} & \textbf{Level} \\ \hline
        Controllability over adaptations & 1 & Limited control over adaptations. \\ 
                                         & 2 & Extensive control over adaptations. \\ 
        Involvement of caregivers        & 1 & Caregivers have no access to the app.  \\ 
                                         & 2 & Caregivers have limited access, such as receiving basic updates or reminders. \\ 
                                         & 3 & Caregivers are actively involved and can jointly utilise the application.\\ 
        Usability of the app             & 1 & Unpredictable and requiring users to learn it due to adaptations. \\
                                         & 2 & Easy to use with a familiar interface. \\ 
        Function usage patterns          & 1 & Less than once a week.\\ 
                                         & 2 & Several times a day. \\ 
        Frequency of adaptations         & 1 & Adapt (only) once when user logs in. \\
                                         & 2 & Adapt weekly. \\ 
                                         & 3 & Adapt monthly.\\
        Granularity of adaptations       & 1 & Adaptations are broad and impact the app significantly. \\
                                         & 2 & Adaptations are specific, affecting small aspects of the app. \\ \hline
    \end{tabular}}
    \label{tab:attributes_levelsnew}
\end{table} 
In a DCE, \textbf{attributes} are key features of a product, service, or system presented to participants as variables influencing their choices, and are typically assigned multiple levels \cite{klojgaard2012designing}. The attributes are extracted from two primary sources: (i) a SLR concerning the current state of the art of AUIs in mHealth applications \cite{wang2024adaptive}, and (ii) the findings of a socio-technical grounded theory analysis that scrutinized data derived from focus groups and interview sessions, focused on examining perceptions regarding the implementation of AUIs in mHealth applications \cite{wang2024adaptiveicse}. The first author identified seven potential attributes from the sources, selecting the six most significant for the DCE survey, which asks participants to assess their importance.

\begin{figure}[h!]
    \centering
    \includegraphics[width=1\linewidth]{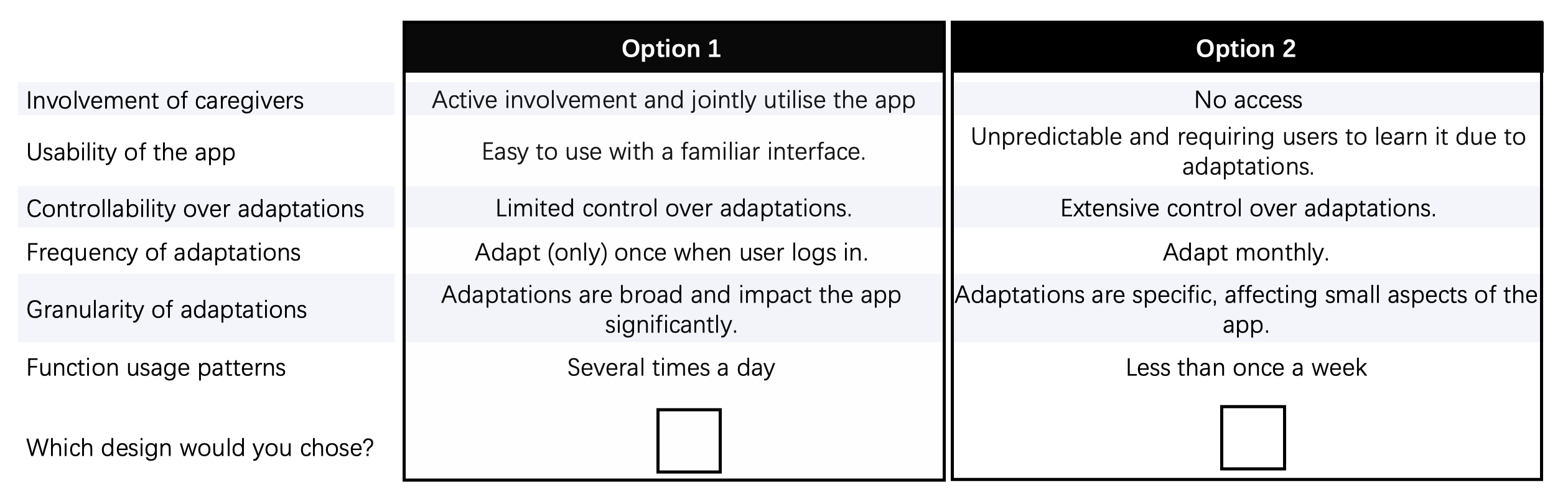}
    \caption{Example question of the DCE}
    \label{fig:discreteexample}
\end{figure}

\begin{enumerate}[leftmargin=20pt, topsep=0pt, itemsep=0pt]
    \item \textbf{Controllability over adaptations}: assesses the level of control users have over how the app adjusts to their needs;
    \item \textbf{Involvement of caregivers}: describes the role caregivers play in supporting daily use of the app;
    \item \textbf{App usability}: refers to how intuitive and user-friendly the app is, even with its adaptive features;
    \item \textbf{Function usage patterns}: reflect how the app adapts to users based on the frequency with which specific functions are used;
    \item \textbf{Adaptation frequency}: examines how often the app’s features change to align with user preferences, ranging from infrequent updates to regular adjustments;
    \item \textbf{Adaptation granularity}: highlights the scale of changes, whether they involve broad adjustments that affect multiple areas of the app or smaller, targeted modifications.
\end{enumerate}
Table \ref{tab:attributes_levelsnew} outlines the various levels for each attribute. An example DCE choice question set for attributes is shown in Figure \ref{fig:discreteexample}. Participants are presented with a list of questions in this format and are requested to select one of the combined choice sets from the two options provided.

\subsection{Survey development}\label{sec:att}
To study how the attributes in Section \ref{sec: attribute} affect user preferences, we used a \textbf{factorial design} to systematically integrate these levels and evaluate their impact on decision-making \cite{hensher2015applied}. The factorial design comprising six attributes, in which four attributes possess two levels each and two attributes present three levels, results in a full factorial design encompassing 144 ($2^4 \times 3^2 = 144$) potential choice sets \cite{hensher2015applied}. However, it is impractical to ask participants to make 72 discrete choices due to time and complexity constraints. Thus, we followed established practice and employed a fractional factorial design, reducing the survey to a subset of choice sets \cite{soekhai2019discrete,bridges2011conjoint,weber2021step}. A \textbf{fractional factorial design} selects a smaller subset of combinations from the full factorial set, effectively capturing key effects. Based on D-optimality criteria, our resulting fractional factorial design consists of 18 choice sets \cite{aizaki2008design}, converted to \textbf{nine} questions, each offering two choice sets from which respondents need to select. We \textbf{\textit{randomized }}both the \textit{order of the attributes} and the \textit{sequence of questions} presented to the participants when entering the survey to minimize order effects, such as consistently prioritizing earlier attributes or questions \cite{bridges2011conjoint}. 

To assess the quality and \textbf{internal validity} of the DCE survey \cite{johnson2019internal}, we added a randomly generated \textbf{tenth} question, along with its set of choices, which mirrored the previous question as one \textit{attention check }question. This facilitates the evaluation of whether the responses exhibited \textit{consistency}, demonstrated by the selection of the same adaptation design in both equivalent questions. Although internal validity checks assess data quality, inconsistent responses to repeated questions align with random utility theory \cite{hess2018revisiting}. To avoid bias or lower statistical efficiency, we \textit{\textbf{do not remove}} participants who fail the validity check, as advised by \citet{lancsar2006deleting}. We provide information on internal validity to allow the reader to assess possible biases, but responses to the random tenth question are excluded from the analysis \cite{szinay2021understanding}. Figure \ref{fig:discreteexample} presents a sample question from the DCE survey, asking participants to choose one of the two presented combined choice sets. We used statistical language R \footnote{https://www.r-project.org} and Qualtrics \footnote{https://www.qualtrics.com} to design the DCE survey.  

\subsection{Piloting and survey adjustments}
The pilot study of our DCE survey involved four participants: two experts specializing in statistics and choice modeling, an experienced academic staff member in the field of survey research, and an individual representative of the target demographic, notably possessing both hypertension and lung cancer. The participants filled out the survey, expressed their opinions, and recorded the time it took them to finish, confirming it was within the 20-minute limit. All participants found the DCE survey engaging and relevant but expressed confusion about \textit{unclear terminology} and \textit{how to respond}. To address these issues, we made the following revisions: (1) simplified the survey language by using clearer, more intuitive terminology; (2) added one comprehension check questions to ensure participants fully understood each attribute, addressing concerns about potential ambiguity; (3) improved survey instructions by including example questions to better guide participants through the response process; and (4) added one more attention check question to ensure thoughtful responses. 

The survey consists of three sections that guide the participants through the study step by step.  Section A begins with an introduction explaining the purpose of the study. It also includes a screening step to confirm participant eligibility (i.e., experience with mHealth apps and a chronic disease), followed by demographic questions such as age, gender, and other relevant background details. Section B introduces participants to the concept of AUI in mHealth applications, explaining their relevance to the study. This section also provides an overview of the DCE, including a sample question and a description of the attributes under investigation. To ensure that participants understand these concepts before proceeding, a comprehension check is included at the end of this section, allowing up to three attempts. The participants then complete ten DCE choice tasks, with the tenth question serving as an attention check to assess the consistency of the internal response. Section C concludes with five questions on participants’ health conditions and their attitudes towards managing health challenges, along with a second attention check to ensure continued engagement with the survey content. A complete version of the survey instrument is available in the supplementary materials for reference\footnote{\url{https://doi.org/10.5281/zenodo.17555183}}.
In total, the survey includes three quality checks: one comprehension check (Section B), one attention check within the DCE (Section B), and one final attention check (Section C).

\subsection{Data collection}
The data collection process for our DCE survey was conducted in two stages. \textbf{Stage One} involved non-probability sampling through community and organizational outreach within Australia. The survey was distributed online through social media platforms and distributed through national and state-level non-governmental health organizations, including the Stroke Foundation, Victoria Diabetes, Dementia Australia and Kidney Health Australia. In addition, physical posters were displayed in private clinics and select hospitals. Collaboration with JOIN US \footnote{https://www.joinus.org.au} facilitated the promotion of the study to individuals with specific health conditions. A total of \textit{122} responses were collected through these channels during stage one. 

In \textbf{stage two}, we employed a purposive sampling strategy through Prolific\footnote{https://www.prolific.com/} by employing a custom screener to refine the pool of participants to individuals with chronic diseases. This change was essential to \textit{increase the sample size} and improve \textit{diversity in participant demographics} since all stage one participants were from Australia. We excluded participants with discrepancies between their screening and prescreening responses on Prolific to identify falsified answers.A total of \textit{64} responses were collected through Prolific during stage two. Overall, we collected \textit{186} responses. To determine the required sample size for a DCE, \citet{orme1998sample} proposes a commonly used rule of thumb estimating a minimum of 94 respondents for our survey design, while research suggests that response precision improves significantly with around 150 participants \cite{johnson2013constructing}, despite the challenges in calculating precise sample sizes for DCEs. Based on this guideline, we set 150 as our benchmark and successfully exceeded this target, ensuring robust and reliable findings.

\subsection{Data analysis}\label{sec:dataanalysis}
Individual characteristics were explored descriptively. To analyze the data collected from the DCE, we used statistical models designed to understand how participants make trade-offs between different features. These models help quantify the importance of each attribute by examining the choices that participants made in multiple scenarios. A \textit{mixed logit model} was used as this modeling approach accommodates multiple
observations per participant and does not require the assumption of the independence of irrelevant alternatives \cite{greene2003latent}. The analysis was performed in R \footnote{https://www.r-project.org}, specifically using the \textit{mlogit} package based on 1000 Halton draws to ensure stable coefficient estimates. A \textit{univariate regression analysis} is conducted to examine the independent effect of the participant's characteristics on choices, which helps to identify the significant factors that affect preference. Following this, a \textit{multivariate regression model} is fitted, incorporating all attributes and participant characteristic factors identified in the univariate analysis, to evaluate the joint effects of multiple factors while controlling for confounding and capturing the relative importance of each. 

Subsequently, an analysis specific to subgroups was conducted to facilitate a direct comparison of preferences across predefined cohorts. The standardized \textit{Relative Attribute Importance (RAI}) was used to elucidate the relative contribution of various attributes in distinct subgroups. A \textit{Marginal Rate of Substitution (MRS)} analysis was conducted to evaluate the trade-offs between attributes by estimating a mixed logit model to derive attribute coefficients. The \textit{\textbf{usability} }of the app is chosen as a reference attribute, given its high significance and its crucial role in the study. Subgroup analyzes are performed to capture heterogeneity in preferences between groups, with separate models estimated for each subgroup. MRS values are normalized, ensuring that the reference attribute had a relative utility of 100\% for ease of comparison. The study examined the influence of seven predetermined participant characteristics on response data. These characteristics were: nationality (Australia or Non-Australia); age; gender; educational attainment (below Bachelor's degree or above Bachelor's degree); health status (Poor (including  fair) or Good); and stress coping mechanism (positive (including mixed feelings) or negative).
% The effects of seven pre-specified participant characteristics on
% response data were investigated. The characteristics were: country
% (Spain or the UK); age; gender; education (categorised as educated to
% age 16, educated to between 16 and 18, and educated beyond age 18);
% user of wearable technology (yes or no); and acceptance of technology
% (UTAUT2) score. Each characteristic was investigated separately and an
% improvement in Akaike's Information Criteria (AIC) was the criterion for
% a significant difference in preferences by subgroup. 

\section{Results}\label{sec:results}
\color{darkgreen}

This section presents the findings of the DCE, including the' overall preferences for adaptive mHealth features, subgroup differences based on demographic and psychological factors, and the trade-offs users are willing to make between competing design attributes.

\color{black}
\subsection{Participants demographic}
\begin{table}[!hb]
\renewcommand{\arraystretch}{0.85}
\centering
\caption{DCE survey participants demographics information (n=186)}
\label{tab:demographics}
    \resizebox{1\textwidth}{!}{%
\begin{tabular}{p{125mm}p{129mm}}
\hline
\toprule
    \begin{tabular}[t]{p{75mm}p{4mm}l}
    \textbf{Demographics} & \textbf{\#}  & \textbf{\% of Participants}  \\ \hline
    \multicolumn{3}{l}{\textit{\textbf{Gender}}}\\ \hline
        Female & 112 & \mybarhhigh{18.7}{60\%} \\
        Male & 74 & \mybarhhigh{12.3}{40\%} \\
        \multicolumn{3}{l}{\textit{\textbf{Age}}}\\ \hline
        18-44 & 93 & \mybarhhigh{15.5}{50\%} \\
        45-84 & 93 & \mybarhhigh{15.5}{50\%} \\ 
    \multicolumn{3}{l}{\textit{\textbf{Ethnicity simplified}}}\\ \hline    
    White & 153 & \mybarhhigh{25.50}{82\%} \\
    Multicultural & 13 & \mybarhhigh{2.17}{7\%} \\
    Black or African American & 12 & \mybarhhigh{2.00}{6\%} \\
    \multicolumn{3}{p{95mm}}{Asian 3\%; Aboriginal or Torres Strait Islander 1\% }\\
    \multicolumn{3}{l}{\textit{\textbf{Health conditions}}}\\ \hline   
    Very Poor & 10 & \mybarhhigh{1.67}{5\%} \\
    Poor & 29 & \mybarhhigh{4.83}{16\%} \\
    Fair & 76 & \mybarhhigh{12.67}{41\%} \\
    Good & 60 & \mybarhhigh{10.00}{32\%} \\
    Very Good & 11 & \mybarhhigh{1.83}{6\%} \\
      \multicolumn{3}{l}{\textit{\textbf{Coping menanism towards stress}}}\\ \hline 
      Avoid dealing with the stress & 13 & \mybarhhigh{2.17}{7\%} \\
       Avoid it but try to stay positive & 34 & \mybarhhigh{5.67}{18\%} \\
        A mix of positive thinking and avoidance & 61 & \mybarhhigh{10.17}{33\%} \\
        Stay positive and try to handle the stress & 57 & \mybarhhigh{9.50}{31\%} \\
        Actively manage the stress  & 21 & \mybarhhigh{3.50}{11\%} \\

    \end{tabular}
&
\begin{tabular}[t]{p{79mm}p{4mm}l}
    \textbf{Demographics} & \textbf{\#}  & \textbf{\% of Participants}  \\ \hline
    \multicolumn{3}{l}{\textit{\textbf{Country of residence}}}\\ \hline
    Australia & 123 & \mybarhhigh{20.5}{66\%} \\
    United States of America & 13 & \mybarhhigh{2.2}{7\%} \\
    Canada & 12 & \mybarhhigh{2.0}{6\%} \\
    Poland & 6 & \mybarhhigh{1.0}{3\%} \\
    Portugal & 6 & \mybarhhigh{1.0}{3\%} \\
    Mexico & 5 & \mybarhhigh{0.8}{3\%} \\
    United Kingdom of Great Britain & 5 & \mybarhhigh{0.8}{3\%} \\
    \multicolumn{3}{p{125mm}}{Chile, Italy and Spain 2\% each; Hungary,  Kenya, Germany, Greece and Sweden 1\% each}\\

    \multicolumn{3}{l}{\textit{\textbf{Education}}}\\ \hline
    Less than Bachelor degree  & 105 & \mybarhhigh{17.50}{56\%} \\
    Bachelor’s degree & 51 & \mybarhhigh{8.50}{27\%} \\
    Master’s degree & 22 & \mybarhhigh{3.67}{12\%} \\
    Doctorate degree & 8 & \mybarhhigh{1.33}{4\%} \\ 

    \multicolumn{3}{l}{\textit{\textbf{Categories of chronic disease**}}}\\ \hline
    Cardiometabolic and Endocrine conditions & 116 & \mybarhhigh{19.3}{62\%} \\
Respiratory conditions & 54 & \mybarhhigh{9.0}{29\%} \\
Immune-related conditions & 22 & \mybarhhigh{3.7}{12\%} \\
Neurological and Mental Health & 41 & \mybarhhigh{6.8}{22\%} \\
Chronic Pain and Musculoskeletal & 34 & \mybarhhigh{5.7}{18\%} \\
Gastrointestinal conditions & 13 & \mybarhhigh{2.2}{7\%} \\
Other conditions & 17 & \mybarhhigh{2.8}{9\%} \\
    
    \end{tabular} \\
        \hline  \multicolumn{2}{p{210mm}}{* Some categories do not add up to 100\% due to rounding.}\\
        \multicolumn{2}{p{210mm}}{** This does not added up to 100\%, because some participants have multiple chronic diseases.}\\

\end{tabular}}
\end{table}

Table \ref{tab:demographics} presents the demographic and key characteristics of our DCE study's 186 participants, who were predominantly women (60\%) and mainly residing in Australia (66\%). Our participants ranged in age, with the largest group aged 65–74 (24\%), followed by those aged 35–44 (20\%) and 25–34 (19\%). The majority identified as White (82\%), with smaller proportions identifying as Multicultural (7\%), Black or African American (6\%), Asian (3\%) or Aboriginal / Torres Strait Islander (1\%). Most of the participants (56\%) did not have a bachelor's degree, while 27\% had a bachelor's degree and 16\% had advanced degrees. Regarding health, 41\% rated their health as fair, while 32\% rated it good, and the remaining categories ranged from poor (16\%) to very poor (5\%). With 62\% of the people reporting cardiometabolic and endocrine disorders, respiratory conditions (29\%) and neurological or mental health problems (22\%). A mix of positive thinking and avoidance is the most common (33\%), followed by staying positive and managing stress (31\%). This diverse sample provides broad information on the demographic and health characteristics of the population.

\subsection{Establishing initial preferences}
\begin{table}[t] 
\renewcommand{\arraystretch}{1}
\centering 
  \caption{Estimated user preferences for adaptive mHealth features based on mixed logit model} 
  \label{tab:main} 
  \resizebox{0.95\textwidth}{!}{%
\begin{tabular}
{lcccccc} 
\hline
\textbf{Attributes} & \textbf{Coefficient} & \textbf{Std\_Error} & \textbf{z\_value }& \textbf{CI\_Lower\textsuperscript{c}} & \textbf{CI\_Upper}\textsuperscript{c}& \textbf{SD}\\
\hline

Controllability over adaptations & $0.279$\textsuperscript{a} & $0.099$ & $2.820$ & $0.085$ & $0.473$ & $0.763$\\ 
Involvement of caregivers & $$-$0.330$ \textsuperscript{a}& $0.081$ & $$-$4.074$ & $$-$0.489$ & $$-$0.171$ & $2.926$\\ 

Usability of the app & $0.704$ \textsuperscript{a}& $0.064$ & $11.079$ & $0.579$ & $0.829$ & $3.137$\\ 

Function usage patterns & $$-$0.229$\textsuperscript{a} & $0.079$ & $$-$2.899$ & $$-$0.383$ & $$-$0.074$ & $2.730$\\ 

Frequency of adaptations & $0.263$\textsuperscript{a} & $0.088$ & $3.007$ & $0.092$ & $0.435$& $3.356$ \\ 

Granularity of adaptations & $0.221$\textsuperscript{a} & $0.102$ & $2.162$ & $0.021$ & $0.421$ & $3.057$\\ 

Log likelihood & -1051.5\textsuperscript{b} &  &  &  &\\ 
McFadden’s Pseudo-R² & 0.113&  &  &  &\\ 
AIC/BIC & 2119.865/2184.946 &  &  &  &\\ 
\hline 
 \multicolumn{7}{p{200mm}}{\small \textbf{a}.
Significant at 5\% level. The table represents beta coefficients and CIs from mixed logit regression. The regression coefficients for each attribute level
represent the mean part-worth utility of that attribute level in the respondent sample. A positive value denotes utility/satisfaction, and a negative value
denotes disutility/dissatisfaction. }\\
 \multicolumn{7}{p{200mm}}{\small  \textbf{b}. The mixed logit model was estimated using maximum simulated likelihood with 1,000 Halton draws. }\\
  \multicolumn{7}{p{200mm}}{\small  \textbf{c}. If the CI does not include zero, the effect is statistically significant. A narrow CI suggests a precise estimate, whereas a wider CI indicates more variability in responses.}\\
\end{tabular} }
\end{table} 
\color{darkgreen}
This subsection presents the main effects from the mixed logit model to identify overall user preferences for adaptive mHealth features, followed by subgroup analyses that examine how these preferences vary across health condition, gender and coping mechanism.
\color{black} In the discrete choice analysis, all attributes are statistically different from 0, indicating their importance in shaping user preferences for adaptation design. Table \ref{tab:main} presents the estimated coefficients, standard errors, z-values, 95\% confidence intervals (CI) and standard deviations (SD) of the random parameters of the mixed logit model. The mixed logit model was estimated using the maximum simulated likelihood with 1,000 Halton draws. The final log-likelihood was –1051.5, with a McFadden’s pseudo-R² of 0.113, indicating a moderate model fit. The AIC and BIC values were 2119.865 and 2184.946, respectively. The standard deviations of the random coefficients reveal considerable heterogeneity in preferences between individuals. This suggests that participants vary substantially in how they value these features, which justifyes the use of a random-parameter approach to better capture individual-level variation. The regression coefficients for each attribute level represent the mean part-worth utility of that attribute level in the respondent sample. A positive value denotes utility/satisfaction and a negative value denotes disutility/dissatisfaction. The usability of the app is the most important factor influencing user preferences ($\beta=0.704$, 95\% CI 0.579--0.829), highlighting the critical role of maintaining usability while adapting the UI. This is followed by controllability over adaptations ($\beta=0.279$, 95\% CI 0.085--0.473), frequency of adaptations ($\beta=0.263$, 95\% CI 0.092--0.435) and granularity of adaptations ($\beta=0.221$, 95\% CI 0.021--0.421), reflecting the value users place on the flexibility of adaptation design. In contrast, the involvement of caregivers ($\beta=-0.33$, 95\% CI $-0.489$ to $-0.171$) and patterns of function usage ($\beta=-0.229$, 95\% CI $-0.383$ to $-0.074$) posed significant obstacles to the adoption of the adaptation design. This suggests that users may be averse to adaptations occurring in features they frequently use, while the involvement of caregivers may lead to a perceived redundancy in adaptations, as users can obtain assistance in utilizing the application from their caregivers. In our DCE, due to the innovative nature of research concerning the adoption of adaptive designs in mHealth technology and the absence of empirical evidence indicating possible attribute interactions, we opted to focus solely on the main effects.

\subsubsection{Differences in the preferences of adaptation design: Subgroup by health condition, gender and coping mechanism}\label{subgroupRAI}
\begin{figure}[ht]
    \centering
    \includegraphics[width=0.9\linewidth]{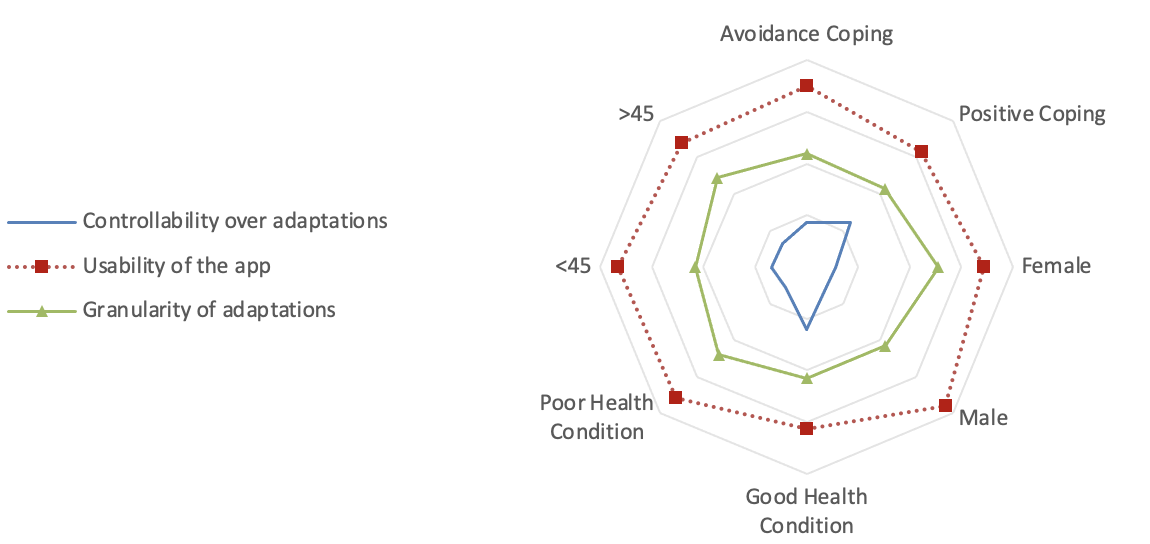}
    \caption{Radar chart for subgroup comparison. This radar chart illustrates the Relative Attribute Importance (RAI) for three key attributes—controllability over adaptations, usability of the app, and granularity of adaptations—across subgroups based on gender, health condition, age, and coping mechanisms.  Lines represent the relative weight each subgroup placed on each attribute. Values are point estimates for descriptive comparison only; no statistical tests were conducted.}
    \label{fig:RAI}
\end{figure}

Following \textit{multivariate analysis} involving country, gender, health condition, and coping mechanism, the significance of the country variable decreased. We selected gender, health condition, age, and coping mechanism as key moderators for our subgroup analysis. We are using RAI to understand the relative contribution of different attributes among different subgroups, as detailed in Section \ref{sec:dataanalysis}, the comparison of subgroups of RAI can be checked in Figure \ref{fig:RAI}. RAI scores reflect how important each attribute is in shaping participants' choices. These RAI values are derived from normalized coefficients and serve as descriptive indicators of preference patterns. Higher scores mean greater influence, while lower scores imply less impact. The granularity of adaptations holds greater importance for women (RAI: 0.399) and older age (RAI: 0.388), those with negative coping styles (RAI: 0.340), and participants in poor health (RAI: 0.370), compared to men and other subgroups. In contrast, controllability over adaptations consistently ranks as the least important attribute across all groups, with minimal importance among men (RAI: 0.008), women (RAI: 0.014) and participants in poor health (RAI: 0.013), although it holds slightly higher value for individuals with positive coping styles (RAI: 0.142) and good health (RAI: 0.142). In general, usability emerges as the paramount attribute, being prioritized most highly by men (RAI: 0.661). It is also more appreciated by people with negative coping mechanisms (RAI: 0.589) compared to those with positive coping strategies (RAI: 0.529), and it is ranked higher by people with poor health (RAI: 0.617) than by those in good health (RAI: 0.529). Because these subgroup analyses are based on point estimates, confidence intervals for RAI values were not computed, and no formal significance testing was performed. As such, subgroup comparisons should be interpreted as descriptive patterns rather than statistically confirmed differences.

\subsubsection{Inconsistent Responses}
A demographic subset of 17 participants (constituting 9.1\% of the total sample, 17 out of 186) exhibited inconsistencies in their responses to the tenth question, as detailed in Section \ref{sec:att}. The predominant portion of these discrepancies was observed among individuals between the ages of 18 and 34 (n = 12), those without a bachelor's degree (n = 9) and participants who reported being in relatively good health (n=14). An analysis comparing the outcomes of the mixed logit model, including and excluding these 17 inconsistent respondents, indicated similar results; therefore, all participants were retained in the final analysis. Importantly, only the duplicated 10th question was excluded from the main model analysis—none of the respondents were excluded. 
 
\subsection{Trade off between different attributes }
\color{darkgreen}
This subsection examines the trade-offs users are willing to make between usability and other adaptation attributes, using MRS to quantify these preferences. The definition and calculation of MRS are detailed in Section \ref{sec:dataanalysis}. \color{black}Table \ref{tab:tradeoffs} displays the MRS values, where a higher positive MRS indicates a greater willingness to trade usability for that attribute, and a negative MRS suggests the attribute is viewed as less valuable or undesirable compared to usability. We found that participants are highly willing to trade usability for controllability over adaptations (40\%) and frequency of adaptations (37\%), highlighting their importance. The involvement of caregivers (-5\%) and function usage patterns (-0.3\%) are perceived unfavorably, indicating that participants regard these factors as less desirable or potentially counterproductive in terms of usability. However, the slight negative valuation of these attributes indicates that they have a relative \textit{importance that is \textbf{closer}} to usability. 

We also performed a subgroup analysis by age, gender, health condition, and coping mechanism based on our previous analysis in Section \ref{subgroupRAI}. Subgroup MRS values were obtained by normalizing point estimates from subgroup-specific models relative to the usability attribute. The confidence intervals for the MRS values were not computed and no formal statistical tests were conducted; therefore, all subgroup comparisons should be interpreted as descriptive rather than inferential. Women prioritized infrequent adaptations (45\%) and controllability (24\%), compared to their male counterparts. Male participants would require an increase of 10\% in usability to compensate for an increase in caregiver involvement. This suggests that caregiver involvement is viewed as a detractor or a feature that reduces the perceived value of the system for men. Young participants value more controllability compared to their older counterparts (54\% versus 25\%). In contrast, older participants prioritize infrequent adaptations (49\%) and small-scale adaptations (35\%). Those in good health prioritize controllability over adaptations (57\%) more than those in poor health (33\%), reflecting a stronger desire for autonomy. Participants in poor health place greater importance on the granularity of adaptations (36\% versus 22\%), emphasizing the need for precision and small-scale changes, while also valuing caregiver involvement (12\%) and less frequent adaptations (4\%). Different approaches to coping led to varied preferences. Those who used positive coping strategies placed a high priority on controllability (85\%), whereas they assigned less significance to infrequent change (28\%). In contrast, participants with negative coping strategies preferred infrequent change (44\%) and smaller changes (36\%) over controllability (19\%). Participants with a positive coping attitude are more open to compromises, allowing caregivers to utilize the app (19\%), whereas those with a negative coping style tend to ignore their health conditions and resist involving the caregiver in the app's usage (-17\%).
 
\begin{table}[!htbp]
\centering
\renewcommand{\arraystretch}{1.2}
\caption{Percentage of usability of the app participants are willing to trade, across the entire sample, and split by gender, age, health condition, and coping mechanism, separately\textsuperscript{a} }
\label{tab:tradeoffs}
\resizebox{1\textwidth}{!}{%
\begin{tabular}{lcccccccccc}
\hline
\textbf{Attributes} & \textbf{All} & \textbf{Man} & \textbf{Woman} & \textbf{Younger} & \textbf{Older} & \textbf{Goodhealth} & \textbf{Poorhealth} & \textbf{Positivecop} & \textbf{Negativecop} \\ 

\hline
Controllability over adaptations & \faArrowUp$40\%$ & \faArrowUp$24\%$ & \faArrowUp $55\%$ & \faArrowUp $54\%$ & \faArrowUp $25\%$ & \faArrowUp $57\%$ & \faArrowUp $33\%$ & \faArrowUp $85\%$ & \faArrowUp $19\%$ \\ 
Involvement of caregivers & \faArrowDown$$-$5\%$ & \faArrowDown$$-$10\%$ & \faArrowDown$$-$0.1\%$ & \faArrowDown$$-$6\%$ & \faArrowDown$$-$3\%$ & \faArrowDown$$-$14\%$  & \faArrowUp $12$\%& \faArrowUp $19\%$ & \faArrowDown$$-$17\%$ \\ 
Function usage patterns & \faArrowDown$$-$0.3\%$ & \faArrowUp $1\%$ & \faArrowDown$$-$2\%$ & \faArrowDown$$-$1.8\%$ & \faArrowUp $1.8\%$ & \faArrowDown$$-$1.2\%$ & \faArrowUp $4\%$ & \faArrowUp$0.2\%$ & \faArrowDown$$-$1\%$ \\ 
Frequency of adaptations & \faArrowUp $37\%$ & \faArrowUp $31\%$ & \faArrowUp $45\%$ & \faArrowUp $26\%$ & \faArrowUp $49\%$ & \faArrowUp $41\%$ & \faArrowUp $39\%$ & \faArrowUp $28\%$ & \faArrowUp $44\%$ \\ 
Granularity of adaptations & \faArrowUp $31\%$ & \faArrowUp $31\%$ & \faArrowUp $32\%$ & \faArrowUp $28\%$ & \faArrowUp $35\%$ & \faArrowUp $22\%$ & \faArrowUp $36\%$ & \faArrowUp $22\%$ & \faArrowUp $36\%$ \\ 
\hline
 \multicolumn{10}{p{260mm}}{ \textbf{a}.
The table represents the Marginal Rate of Substitution (MRS) expressed as percentages, derived from the mixed logit regression model. The percentages indicate the degree to which participants are willing to trade off usability (the reference attribute) for improvements in other attributes. 
\begin{itemize}[left=-8pt,topsep=0pt, itemsep=0pt]
    \renewcommand\labelitemi{}
    \item \faArrowUp\, \textbf{Positive values}: Indicate that participants are willing to compromise usability to prioritize improvements in that attribute. Higher percentages reflect greater willingness to trade usability for improvements in the respective attribute. 
    \item \faArrowDown\, \textbf{Negative values}: Indicate that participants are willing to sacrifice that attribute to preserve usability, implying lower desirability or priority of the attribute relative to usability.
    \end{itemize}}\\

\end{tabular}}
\end{table}

\section{Discussion}\label{sec:discussion}

% Relative Attribute Importance (RAI) scores complemented the MRS findings, highlighting the proportional contribution of each attribute to decision-making while emphasizing subgroup-specific differences.

\subsection{Implications for mHealth Developers}

\begin{figure}
    \centering
    \includegraphics[width=1\linewidth]{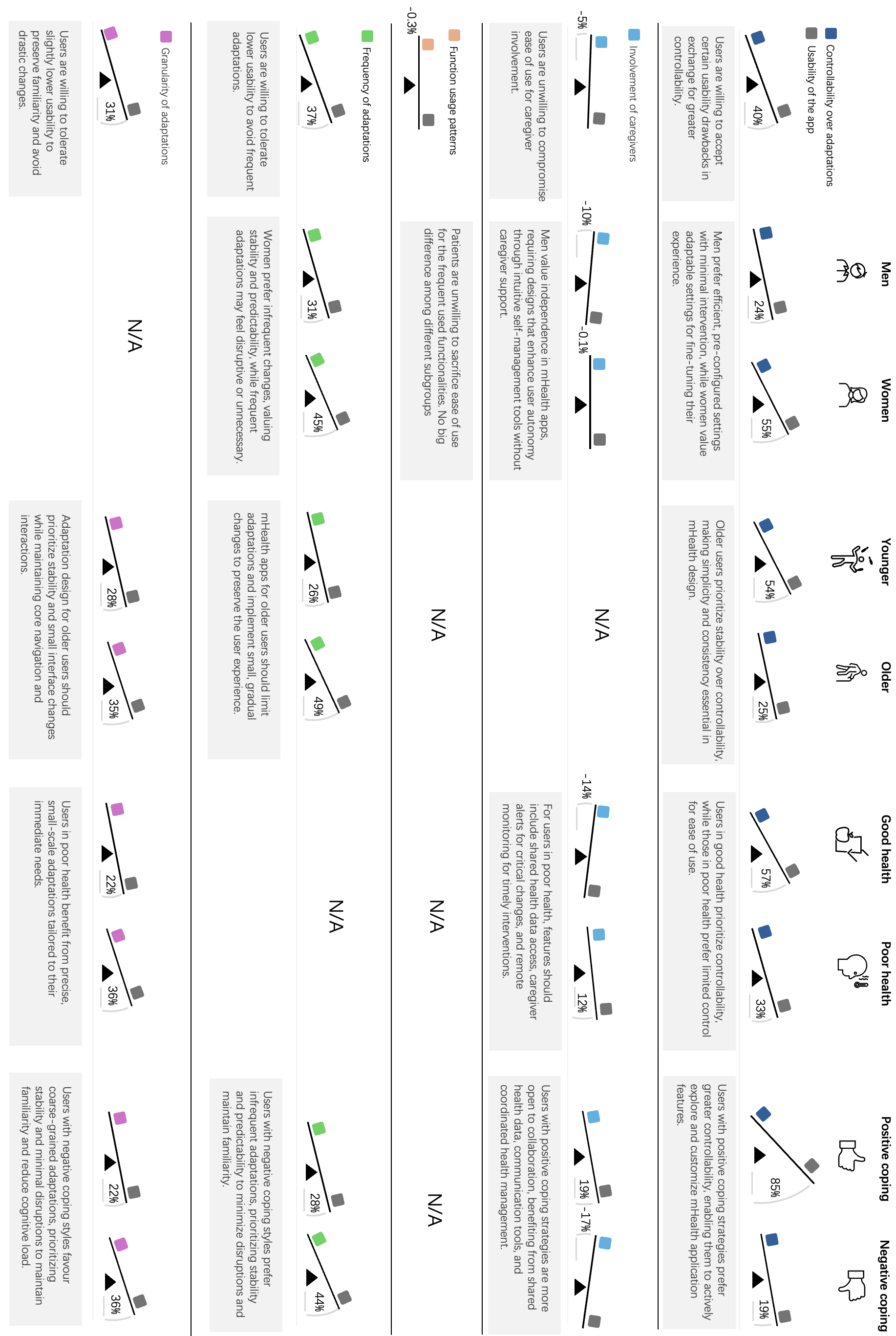}
    \caption{MRS relative to usability and its implications (N/A indicates no significant difference in preference for that attribute across subgroups).}
    \label{fig:tradeoff}
\end{figure}

Our DCE study explored participants' preferences for various attributes in mHealth app adaptation design and their willingness to make trade-offs, focusing on different subgroups such as gender, coping strategies, and health conditions. Figure \ref{fig:tradeoff} visually illustrates the trade-offs between different attributes and delineates how these trade-offs vary between different subgroups. The findings highlight app \textbf{usability} as the most critical attribute across all subgroups, underscoring its pivotal role in users' decision-making processes regarding adaptation design. However, the study also reveals that providing users with \textbf{controllability over adaptations}, such as the ability to remove specific features, revert changes, or adjust adaptation levels, can mitigate potential usability trade-offs. User-controlled adaptation fosters a sense of ownership and agency, leading to increased engagement and motivation to utilize the system effectively \cite{sundar2010personalization,sundar2008self}.  \color{darkblue}While caregiver involvement can offer benefits, the observed negative attitudes indicate that adaptation mechanisms should be carefully designed to balance control and support. \color{black}

Incorporating \textbf{caregivers} into mHealth systems can influence user preferences for app adaptations, as patients can delegate certain responsibilities to caregivers. However, involving caregivers can also create a negative attitude toward adaptations. For example, apps like \textit{Health2Sync}\footnote{https://www.health2sync.com/} and \textit{LibreLinkUp}\footnote{https://www.librelinkup.com/} include caregivers in chronic disease management but limit their role to monitoring, without allowing them to control or customize app features for patients. This suggests that while caregiver participation can be beneficial, practitioners should carefully design adaptation mechanisms to maximize benefits.  
 In particular, \textit{permit caregivers to \textbf{aid in distinct tasks}} (such as personalizing reminders or monitoring health information) instead of granting patients sole control over the application's capabilities. Ensure that adaptations do not degrade the app’s usability, as patients are unwilling to sacrifice ease of use for caregiver involvement.

The \textbf{frequency and criticality} of the app functionalities play a significant role in determining user preferences for adaptations. Users are less inclined to accept adaptations for frequently used or essential functionalities, as these tasks require consistency and reliability. This aligns with the findings of \citet{lavie2010benefits}, who demonstrated that the proportion of routine versus non-routine tasks influences the effectiveness of AUIs. Specifically, adaptations are less desirable for tasks that are performed regularly or are critical to the user’s workflow.  For mHealth applications, where users often engage in a mix of routine and non-routine tasks \cite{bunt2004role}, this insight is particularly relevant. Practitioners should \textit{avoid introducing many adaptations for commonly used or essential functionalities} to maintain user trust and efficiency. Focus on adapting features that are used infrequently or are less critical, where users may be more open to changes. Ensure that adaptations \textit{do not disrupt the usability of core functionalities}, as these are key to user satisfaction and engagement.

Patients prefer adaptations that \textbf{occur infrequently and only when necessary}. Once an adaptation aligns with their capabilities and needs, they see little value in frequent changes \cite{wang2024adaptiveicse}. This suggests that adaptations should focus on essential functionalities and aim to enhance usability without introducing unnecessary disruptions. Importantly, users are willing to tolerate lower usability if it means avoiding frequent adaptations, particularly for less-used features. To create effective adaptive mHealth applications, practitioners should minimize adaptation frequency by designing changes to occur only when absolutely necessary.  Practitioners should also focus on \textit{core functionalities by prioritizing adaptations for essential tasks that significantly improve usability or accessibility}. Additionally, it is crucial to avoid over-adaptation by resisting frequent changes. Frequent adaptations can lead to frustration and reduced trust in the app.

Patients prefer small-scale and incremental adaptations, ensuring that the app remains familiar even after adaptations are implemented. For example, they are more comfortable with adaptations that modify specific components, such as the navigation bar or the order of data entry fields, rather than large-scale overhauls of the interface. \color{darkblue}This finding suggests that familiarity may outweigh usability improvements for some users. \color{black}Practitioners should make \textbf{small, specific adaptations} to app components rather than extensive modifications. Maintain the app's structure and flow to keep it familiar to users, recognizing their preference for familiarity and consistency, even at the expense of some usability improvements.

 \textbf{Summary.} When designing adaptive mHealth apps \textbf{without caregiver involvement}, practitioners should focus on less frequently used functionalities for adaptations, as these are less likely to disrupt the user experience. For example, adaptations could target features like data export settings or advanced analytics, which users interact with infrequently. These adaptations should provide essential assistance while being infrequent and small-scale to avoid overwhelming the user. When \textbf{caregivers are involved in mHealth app usage}, the design should prioritize collaborative functionality while maintaining a balance between patient autonomy and caregiver support. For example, caregivers could have access to monitor health data or set reminders, but patients should retain control over critical app customizations to ensure usability and comfort. For the rest of the factors for adaptation design. The design should follow the recommendations outlined earlier, such as focusing on small-scale, infrequent changes and prioritizing less frequently used functionalities. To address the \textbf{varying frequency of functionality usage}, a practical design approach is to introduce adaptations for less frequently used features during the first login. This allows users to customize these features upfront, ensuring personalization without the need for frequent changes later. For frequently used functionalities, such as daily health tracking or medication reminders, it is better to let users interact with the app for a period of time before offering adaptations.

\subsubsection{Heterogeneous preferences across different user groups}
\textbf{Gender}. Our analysis indicates that while men prioritize usability, women place greater emphasis on adaptability and customization, underscoring the importance of gender-specific design considerations (See Figure \ref{fig:tradeoff}). According to Hofstede’s Masculinity vs. Femininity dimension, masculine cultures tend to favor practical problem-solving strategies, whereas feminine cultures place greater value on emotional support and personalized care \cite{ahmed2024cultural}. These preferences align with the behavioral differences observed in the usage of mobile applications.  For example, women tend to use smartphones for longer periods than men \cite{andone2016age}, and their app-related gratifications are more focused on enjoyment and setting goals for physical activity tracking applications. In contrast, men are more likely to engage in live tracking and sharing of their results \cite{klenk2017gender}. To improve usability for male users, designs should prioritize streamlined interfaces, performance-oriented analytics, and quick-access shortcuts, ensuring efficiency without excessive complexity. In contrast, female users benefit from greater adaptability through personalized settings, context-aware assistance, motivational feedback, and progress-tracking tools that align with their preference for social and emotional engagement. Meanwhile, men place less value on caregiver involvement, so the design of mHealth applications should focus on improving user autonomy by providing intuitive self-management tools, clear guidance for independent use, and seamless access to necessary resources without relying on external support. 

\textbf{Age.} A clear pattern emerged among older individuals, who were more likely to trade off usability for less frequent and small-scale adaptations while placing less emphasis on controllability compared to younger users (See Figure \ref{fig:tradeoff}). This preference may be due to age-related memory decline and limited familiarity with technology, which can increase susceptibility to confusion and cognitive overload \cite{isakovic2016usability, wilson2021barriers, li2021design}. Frequent changes or an abundance of options within an application can feel overwhelming for older users, making stability and simplicity in design critical \cite{garcia2018enabling}. The current mHealth landscape is deficient in user-friendly tools and services, with elderly people particularly facing usability challenges when using mHealth technology \cite{wildenbos2019mobile}. Extensive research has been conducted on designing mHealth applications to better suit elderly users, focusing on features such as simple and consistent navigation \cite{wildenbos2015framework,kalimullah2017influence}, presenting all essential content on one screen and using a simple and familiar language \cite{wildenbos2019mobile,pan2021perception}. Additionally, cultural considerations in UI design have been highlighted as critical for improving usability for older adults \cite{alsswey2020elderly}. \color{darkblue}Based on the observed preferences, mHealth applications for elderly users may benefit from infrequent, small-scale adaptations that modify only specific parts of the interface while preserving familiarity and ease of navigation.\color{black}

\textbf{Coping mechanism.} Patients must continually develop strategies to maintain their physical, emotional, and spiritual health due to the long-lasting nature of chronic disease \cite{bussing2010adaptive}, making coping with the disease an ongoing process. Some individuals cope by ignoring distressing situations through avoidance and denial, while others use positive strategies to alleviate stress and restore normal functioning \cite{folkman1980analysis}.
Coping strategies affect consumers' beliefs about new technology, which subsequently mediate the impact of these strategies on their attitude towards adoption \cite{cui2009consumers}. \color{darkblue}Our findings suggest that coping strategies shape users’ preferences for adaptation design. For individuals with positive coping styles, adaptive mHealth designs can benefit from supporting collaboration through caregiver involvement, such as shared access to health data and communication tools, while also providing customization options that improve user control over app functionality. In contrast, the observed preferences of individuals with negative coping styles indicate a stronger need for simplicity and stability, favoring infrequent and small-scale adaptations that preserve familiarity and minimize disruption (see Figure~\ref{fig:tradeoff}). These patterns suggest that caregiver integration should remain optional, respecting preferences for independent use, while subtle prompts and non-intrusive notifications may help encourage engagement without imposing additional burden.\color{black}

\textbf{Health conditions.} Chronic diseases impose diverse physical, psychological, and mental effects on individuals \cite{harvey2012future, di2019chronic}. Research has shown that changes in health conditions over time for those with chronic diseases significantly affect their use of technology, a crucial subject that remains underexplored \cite{paymal2024good}. Health conditions played a critical role in shaping attribute preferences. Participants in good health prioritize controllability, reflecting a preference for autonomy, while those in poor health emphasize small-scale adaptations and caregiver involvement, indicating a need for support and precision (See Figure \ref{fig:tradeoff}). The design of mHealth technology frequently does not account for the fluctuation of mHealth conditions \cite{paymal2024good,mack2022chronically}, and most research tends to focus on a single chronic condition, despite the increasing prevalence of multimorbidity over time \cite{islam2014multimorbidity, sockolow2021integrative}. \color{darkblue}These findings suggest that adaptations should be tailored to accommodate users with diverse health conditions. For individuals in relatively good health, emphasizing controllability—through customization settings, manual override functionalities, and interfaces that allow users to fine‑tune their experiences—may better support autonomy. In contrast, the observed preferences of participants in poorer health indicate greater benefit from simple, small‑scale design modifications and straightforward caregiver access for supervision and assistance. As health conditions can change over time, these results further suggest that mHealth applications should incorporate dynamic adaptation mechanisms that adjust features in response to users’ current health status.\color{black}

\subsection{Implications for researchers}

\subsubsection{Using DCEs to Design mHealth Applications} The growth of mHealth apps has exceeded the evidence of their effectiveness, creating a gap in understanding the optimal design to meet user needs \cite{kumar2013mobile}. Many mHealth interventions are designed around existing healthcare systems, often \textit{without sufficient input from end users} \cite{mccurdie2012mhealth}. This user-centered design shortfall can lead to apps that are hard to use or do not meet users' specific needs, prove ineffective, or, in some cases, lead to adverse outcomes \cite{brown2013assessment, schnall2016user}. In a healthcare system already burdened by suboptimal outcomes and rising costs, premature adoption of untested mHealth technologies risks exacerbating these challenges. Using a DCE can inform the design of mHealth applications by identifying key features that promote user \textbf{uptake and engagement} \cite{jonker2020covid,simblett2023patient}. Unlike traditional methods, the experimental nature of DCEs enables researchers to systematically vary attributes under controlled conditions, providing insight into the marginal effects of attribute changes on user choices \cite{johnson2013constructing}. 

By identifying \textbf{high-priority} features early in the design process, DCEs can help reduce\textit{\textbf{ development costs}} and ensure that \textit{\textbf{resources}} are allocated to features that maximize user engagement and satisfaction. For example, in our DCE study, we found that involving caregivers in the app design significantly influenced patients' perceptions of adaptations, highlighting the importance of considering the interdependence between adaptation features and user roles. DCEs also provide a structured approach to understanding how different \textbf{features interact} and influence user preferences. For instance, in our DCE study, the level of controllability over the adaptations can have cascading effects on the app usability. By quantifying these trade-offs, DCEs enable researchers to make informed design decisions that balance \textit{competing priorities and align with user needs.}

However, while DCEs offer significant advantages, they also present challenges, such as the \textit{complexity of designing choice tasks} and the need for \textit{large sample sizes}. There are existing tools (e.g., Qualtric, Ngene\footnote{https://www.choice-metrics.com/index.html}, SPSS\footnote{https://www.ibm.com/products/spss-statistics}) and papers (e.g., \cite{weber2021step}, \cite{bridges2011conjoint}, \cite{szinay2021understanding}) to teach how to do the DCE step by step. The methodology remains challenging, especially for those without prior experience, as the intricate process of creating realistic choice scenarios, ensuring attribute balance, and analyzing complex data poses significant barriers to entry for novice researchers. Future research could focus on developing tools to simplify the implementation of DCEs for researchers and practitioners, making the methodology more accessible to those without extensive experience. Exploring \textbf{\textit{mixed methods}} that integrate DCEs with qualitative approaches, such as interviews or focus groups, could offer deeper insights into user preferences by allowing researchers to quantify trade-offs and comprehend the contextual factors and personal experiences influencing decisions.

\subsubsection{Identifying and Meeting Non-Functional Requirements (NFRs) } 
\textbf{Non-functional requirements (NFRs)} are global constraints such as performance, security, and availability that software must meet, often across multiple system components \cite{rosa2002processnfl}. These requirements play a critical role in ensuring software quality, user satisfaction, and project success. However, mismanagement of NFRs has been a major cause of many software project failures \cite{ryan2000approach}, often due to their \textit{inherent complexity, interdependencies, and conflicts}. Various methods, such as fuzzy logic \cite{saadatmand2015fuzzy}, ontology-based techniques \cite{mairiza2013conflict,shah2019ontological}, and machine learning or deep learning \cite{alashqar2022studying}, automate the extraction, classification, and prioritization of NFRs from requirements documents. However, these approaches often fail to capture end-user preferences and priorities, limiting their effectiveness in designing user-centered systems. The question of \textit{how users perceive and manage these competing NFRs remains largely unexplored}. For example, in mHealth applications, users may value privacy highly, but may be unwilling to sacrifice usability or performance for improved privacy \cite{hamidi2018should}. DCEs offer a promising framework to address this gap by systematically quantifying user preferences for such trade-offs. By simulating real-world decision-making scenarios, DCEs allow researchers to uncover \textit{\textbf{hidden priorities}} and inform design decisions that align with user needs. In addition, DCEs could address evolving challenges in NFR management, such as sustainability and ethical considerations, ensuring that software systems are not only functional, but also socially responsible.

\subsubsection{How do user preferences differ among various user demographics} 

Understanding how user preferences differ across demographics, including age, health conditions, and mechanisms for managing stress related to diseases, is critical to designing inclusive and effective mHealth solutions. Traditionally, mHealth applications have \textit{focused mainly on younger and healthier audiences} \cite{boulos2011smartphones}, often neglecting the needs of older adults or individuals with chronic diseases. This lack of inclusive design risks \textit{exacerbating health disparities} and \textit{undermining the goal of universal access to healthcare services} \cite{van2022digital}. Insights from DCE can also guide inclusive design practices, ensuring that mHealth apps are accessible and effective for all users. By focusing on the requirements of underrepresented groups, including elderly individuals or those with serious chronic disease, DCEs can work towards diminishing health inequalities and foster fair access to medical services. Beyond mHealth, DCEs have broader applicability in domains such as education, finance, and e-commerce, where understanding user preferences is critical to designing effective systems. For example, in education, DCEs can help prioritize features for personalized learning platforms \cite{marienko2020personalization}, while in e-Commerce, they can inform the design of user-friendly interfaces.

\section{Threats to validity}\label{sec:threat}
 
\textbf{Internal Validity. } First, the study relied on \textit{self-reported data}, which can introduce biases such as social desirability or inaccurate recall \cite{chan2010so}. Although there were efforts to mitigate biases, these may have influenced the’ stated preferences of the participants, and as with all stated preference methods, the results reflect hypothetical choices in a controlled survey context, which may not fully align with their revealed preferences in real-world use. Furthermore, variations in the way the participants interpreted the attributes may have occurred, despite a \textbf{\textit{comprehension check}} carried out before the DCE. We also included a random tenth question that mirrors the previous to assess whether responses showed \textit{ consistency} by choosing the same adaptation design. Furthermore, while the CIs in Table \ref{tab:demographics} provide insight into participant understanding, this variation constrains the strength of our conclusions. An excessive number of choice sets or overly complex attribute combinations in a DCE can result in participant fatigue or cognitive overload, potentially causing random or inconsistent responses. To address this, we limited the survey to a selection of choice sets, implementing a \textit{fractional factorial design} \cite{soekhai2019discrete,bridges2011conjoint}. The sequence in which choice sets or attributes are presented could affect responses \cite{bridges2011conjoint}, and participants could alter their decision-making approach as the experiment progresses, leading to inconsistent responses. To mitigate order effects, we randomly assigned both the order of attributes and the sequence of questions each participant received upon survey entry.

\textbf{External Validity.} The \textit{sample size}, while sufficient for statistical analysis, may not fully represent the diversity of people with chronic diseases, particularly in terms of cultural, socioeconomic, and technological backgrounds. Furthermore, 62\% of the participants had cardiometabolic and endocrine conditions (Table \ref{tab:demographics}), which may limit the generalizability of the findings to populations with other health conditions. Similarly, the participants’ familiarity with technology may have influenced their preferences, restricting broader applicability to less tech-savvy groups. Using Prolific as a recruitment platform could potentially lead to \textit{sampling bias}. Despite Prolific's reputation for a varied set of participants, its users generally possess technological proficiency and a familiarity with online surveys. This might skew the sample towards people who already found mHealth applications easy to use. Although Prolific implements quality control measures, the platform relies on self-reported health information, which can introduce inaccuracies or misclassification in participant eligibility. Furthermore, integrating Prolific's responses with those obtained through alternative recruitment methods could result in \textit{heterogeneity of the sample}, as participants sourced from various channels could vary in their demographic or psychographic characteristics. 

\textbf{Construct Validity.} The DCE methodology, while effective in understanding trade-offs and preferences, simplifies real-world decision-making by \textit{isolating attributes and limiting context} \cite{de2012discrete,soekhai2019discrete}. The \textit{\textbf{simplification and hypothetical nature }}of DCEs may not adequately reflect the intricate requirements of users in adaptive mHealth applications, raising doubts about whether the choices in DCEs align with real-world decisions, known as the intention-behavior gap \cite{ajzen1991theory}. Although participants may assume that they would choose a scenario from a set of options, in real-world contexts, additional factors, such as \textit{privacy concerns} regarding adaptation or \textit{user support} of adaptation \cite{trewin2000configuration, hamidi2018should}, could influence their decisions.

\textbf{Conclusion Validity.}   
Our sample size was sufficient for the primary analyzes, but it limited the ability to explore interactions between demographic factors (e.g., age and health condition) without substantially reducing statistical power. Further stratification would have produced smaller subgroups, increasing the risk of unstable estimates and Type II errors. Future studies with larger and more diverse samples are needed to examine these interactions with greater reliability.

\section{Conclusion}\label{sec:conclusion}\label{sec:conclusion}
\color{darkblue}
This study contributes a data-driven approach to modeling user preferences and trade-offs in adaptive mHealth interface design, using a Discrete Choice Experiment (DCE) with individuals managing chronic diseases. By prioritizing attributes such as usability, controllability over adaptations, frequency of adaptations, and granularity of adaptations, the findings underscore the importance of designing mHealth applications that align with user needs and preferences. This study contributes to the advancement of adaptation design for mHealth applications by offering a robust framework for understanding and addressing competing user requirements. Through a data-driven approach to requirements prioritization, we quantify user preferences and assess the relative importance of key attributes. By analyzing preference heterogeneity, we reveal how preferences vary across demographic and behavioral subgroups, highlighting the trade-offs users are willing to make. These insights inform the design of more inclusive and responsive mHealth applications. Future work could extend this framework to support dynamic personalization strategies and integrate user-driven prioritization into broader requirements engineering processes for adaptive health technologies.

\color{black}

\section*{Acknowledgements}
Wang and Grundy are supported by the ARC Laureate Fellowship FL190100035. Our recruitment was partly supported by Join Us, a National, not-for-profit and disease agnostic research register founded by UNSW and The George Institute. We express our gratitude to all participants who completed our survey and those who provided valuable feedback to help refine it. We express our sincere gratitude to Dawn Mischewski for her invaluable assistance in refining the survey design.

\section*{Declaration of generative AI in scientific writing}
During the writing process of this work, the authors used ChatGPT to improve and rephrase written paragraphs. After using this tool, the authors reviewed and edited the content as needed and take full responsibility for the content of the publication.
\clearpage
\newpage
\appendix

%% If you have bibdatabase file and want bibtex to generate the
%% bibitems, please use
%%

\bibliographystyle{elsarticle-num-names} 
\bibliography{reference}

%TC:endignore
\end{document}